\documentclass[11pt,twoside]{article}
\usepackage{asp2004}
\usepackage{psfig}
\usepackage{epsf}
\usepackage{graphics}
\usepackage{lscape}
\def\edcomment#1{\iffalse\marginpar{\raggedright\sl#1\/}\else\relax\fi}
\reversemarginpar

\begin{document}
\title{HST spectral mapping of V2051~Ophiuchi in a low state}
\author{R. K. Saito, R. Baptista}
\affil{Universidade Federal de Santa Catarina, Brazil}

\begin{abstract}

We report a study of the spectra and structure of the accretion disc
of the V2051 Oph while the star was in an unusual, faint brightness
state during 1996.
\end{abstract}
\thispagestyle{plain}

\section{Observations and data analysis}

V2051 Oph ($P_{orb}=90\,min$) belongs to a small group of ultrashort
period eclipsing dwarf nova together with Z Cha, OY Car and HT Cas.
During 1996, it was seen at a low brightness state ($B\simeq16,2$), in
which mass accretion was reduced to a minimum (Baptista et al. 1998).

High-speed ultraviolet and optical spectroscopy was secured with the
Faint Object Spectrograph onboard the Hubble Space Telescope on 1996
Jan 29 (UV: G400H grating, $\lambda\ 3226-4781$~\AA,~$\Delta\lambda=
1.5$ \AA\ pixel$^{-1}$; optical: G160L grating, $\lambda\
1150-2507$~\AA,~$\Delta\lambda= 3.5$ \AA\ pixel$^{-1}$) at a time
resolution of 3.4\,s.  The runs cover the eclipse cycles 109\,988 (UV)
and 109\,989 (optical), according to the ephemeris of Baptista et
al. (2003).

The out-of-eclipse UV spectra show prominent Ly$\alpha$ $\lambda
1216$, C\,II $\lambda 1323$, Si\,IV $\lambda 1394,1403$ and C\,IV
$\lambda 1549,1551$ emission lines as well as broad absorption bands
possibly due to Fe\,II. The optical spectra show the Balmer continuum
in emission as well as He emission lines and strong Balmer emission
lines.

The UV spectra were divided into 34 narrow passbands (19-60 \AA\ wide
in the continuum and velocity-resolved bins of 2000~km/s or 3000~km/s
for the lines) while the optical spectra were divided into 68
passbands (15-42 \AA\ wide in the continuum and velocity-resolved bins
of 1000~km/s for the lines) and light curves were extracted for each
one.  Maximum-entropy eclipse mapping techniques (Baptista \& Steiner
1993) were used to solve for a map of the disc brightness distribution
and for the flux of an additional uneclipsed component in each band.

\section{Results}

The line centre maps show broad brightness distributions from an
extended and optically thick region, possibly ionized by the boundary
layer. Velocity-resolved H$\gamma$ line light curves show the expected
behavior for the eclipse of gas rotating in prograde sense (rotational
disturbance), with the blue wing ($-1000\,km/s$) being eclipsed
earlier than the red wing ($+1000\,km/s$). Net line emission maps show
that the Balmer lines are in absorption at the disc centre, and that
there is also little contribution from the bright spot at the disc
rim. In contrast with the observed in the line maps, the continuum
maps show pronounced emission from the compact white dwarf and bright
spot sources as well as enhanced emission along the ballistic stream
trajectory of gas stream, an evidence of gas stream overflow.

We sliced the disc into concentric annullar regions of width $0.1\,\rm
R_{L1}$ to compute disc spectra as a function of distance.  Motivated
by the asymmetric emission patterns observed in the eclipse maps, we
extracted spatially resolved spectra for three distinct disc regions:
``front'' (the disc side closest to the secondary star), ``back'' (the
side disc farther away from the secondary star) and ``gas stream'' (a
thin slice around the stream trajectory in each annullar region).

The disc (front and back) spectra show that the Balmer decrement
becomes shallower and the lines are more prominent and narrower with
increasing radii. The ``front'' side is systematically brighter than
the ``back'' side in agreement with the results of Vrielmann et
al. (2002). The difference becomes more pronounced as the radius
increases. The comparison of the ``front'' and ``back'' spectra show
that the latter contain broad absorption bands, possibly due to
Fe\,II, that become more conspicuous in the outer parts of the disc --
suggesting that they arises from absorption along the line of sight in
a vertically extended region above the disc (e.g. Horne et
al. 1994). The spectrum of the central pixel shows a continuum filled
with broad and shallow absorption lines and a Balmer jump in
absorption, and resembles that of a DA white dwarf.

We computed the ratio between the spectrum of the ``gas stream'' and
that of the ``front'' side at the same radius to investigate the
emission along the gas stream trajectory. The spectrum of the gas
stream is clearly distinct from the disc spectrum for a large range of
radii. In contrast to the observed in the disc spectra, the lines
appear in absorption.  The distinct emission along the gas stream
trajectory underscores the evidence of gas stream overflow.  The fact
that the lines appear in absorption in the stream spectra suggest the
presence of matter outside of the orbital plane (e.g. a chomosphere +
wind).

The uneclipsed component is dominated by the Balmer continuum and
strong Balmer emission lines in the optical and by resonant emission
lines in the UV, suggesting that it arises in a vertically extended
region. The strong lines in the optical and UV spectra as well as the
Balmer continuum accounts for $\sim 5-10$ per cent of the total light
at the respective wavelength. The continuum in the uneclipsed spectrum
yield negligible fractional contributions to the total light at the
corresponding wavelengths.

We are fitting LTE atmosphere models to the spatially resolved disc
spectra to derive the radial run of the temperature, surface density,
turbulent mach number and vertical scale height.

\end{document}